\documentclass[aip,prb,lineno,twocolumn,reprint,showpacs,superscriptaddress]{revtex4}
\usepackage{graphicx}
\usepackage{epsfig}
\usepackage{color}
\usepackage{amsmath}
\usepackage{lineno}

\def\dd{\mbox{d}}

\newcommand{\pt}[2]{\frac{\partial{#1}}{\partial{#2}}}
\newcommand{\ptn}[3]{\frac{\partial^{#3}{#1}}{\partial{#2}^{#3}}}

\newcommand{\bfrac}[2]{\left(\frac{#1}{#2}\right)}
\newcommand{\vgap}{\vspace{0.1cm}}
\graphicspath{{./eps/}}

\begin{document}
\bibliographystyle{apsrev}



\title{Charge transport-mediated recruitment of DNA repair enzymes}


\author{Pak-Wing Fok}
\affiliation{Applied and Computational Mathematics, California Institute of Technology, CA 91125}
\affiliation{Dept. of Biomathematics, UCLA CA 90095-1766}
\email[]{pakwing@caltech.edu}

\author{Chin-Lin Guo}
\affiliation{Applied Physics and Bioengineering, California Institute of Technology, CA 91125}

\author{Tom Chou}
\affiliation{Dept. of Biomathematics, UCLA CA 90095-1766}
\affiliation{Dept. of Mathematics, UCLA CA 90095-1766}
\email[]{tomchou@ucla.edu}


\date{\today}

\begin{abstract}
Damaged or mismatched bases in DNA can be repaired by Base Excision
Repair (BER) enzymes that replace the defective base.  
Although the detailed molecular structures of many BER enzymes are
known, how they colocalize to lesions remains unclear. One hypothesis
involves charge transport (CT) along DNA [Yavin, {\it et al.}, PNAS,
{\bf 102}, 3546, (2005)].  In this CT mechanism, electrons are
released by recently adsorbed BER enzymes and travel along the DNA.
The electrons can scatter (by heterogeneities along the DNA) back to
the enzyme, destabilizing and knocking it off the DNA, or, they can be
absorbed by nearby lesions and guanine radicals.  We develop a
stochastic model to describe the electron dynamics, and compute
probabilities of electron capture by guanine radicals and repair
enzymes. We also calculate first passage times of electron return, and
ensemble-average these results over guanine radical distributions.
Our statistical results provide the rules that enable us to perform
implicit-electron Monte-Carlo simulations of repair enzyme binding and
redistribution near lesions.
When lesions are electron absorbing, we show that the CT mechanism
suppresses wasteful buildup of enzymes along intact portions of the
DNA, maximizing enzyme concentration near lesions.
\end{abstract}


\maketitle
\section{Introduction}

The genomes of all living organisms are constantly under attack by
mutagenic agents such as reactive oxygen species and ionizing
radiation. Such processes can damage bases giving rise to localized
lesions in the DNA \cite{BRUNER,NASH} that can lead to harmful
mutations and diseases such as cancer. For example, guanine residues
can be oxidized, generating a radical called 7,8-dihydro-8-oxoguanine,
or oxoG for short. Unlike the non-oxidized form, this radical can pair
with both cytosine and adenine, ultimately giving rise to GC
$\rightarrow$ TA transversion mutations \cite{BRUNER} upon multiple
replications. Lesions can also arise through alkylation, hydration and
deamination.\cite{BRUNER}

One defense mechanism against these mutation processes is the Base
Excision Repair (BER) pathway. BER enzymes recognize and undo damage
to DNA by adsorbing onto the sugar-phosphate backbone, locating the
lesion and excising it.  The biomechanical functions of repair enzymes
have been well established and their 3D structures are known in great
detail.\cite{PARIKH} There are four main types of BER enzyme: DNA
glycosylases, AP-endonucleases, DNA polymerases and DNA ligases. Each
of these enzymes has a different role in the BER family.  For example,
DNA glycosylases initiate the repair pathway, detecting and
recognizing distinct forms of DNA damage while the endonucleases are
responsible for cleaving the sugar-phosphate backbone. Together, these
enzymes maintain the overall integrity of DNA, generally ensuring that
miscoded proteins are kept to a minimum.

\begin{figure}[htbp]
\begin{center}
\includegraphics[width=3.1in]{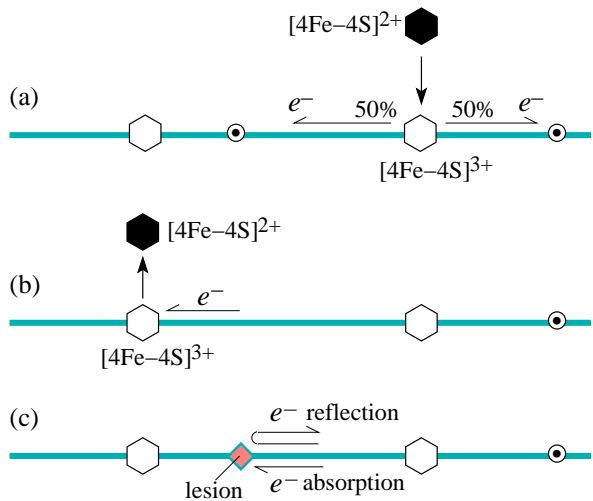}
\end{center} 
\caption{Redox mechanism for repair enzyme interaction based on the
papers by Yavin, {\it et al.}\cite{BARTON} and Boon, {\it et
al.}\cite{BARTON2} (a) A MutY in the $2+$ state (solid hexagon)
adsorbs and oxidizes to the $3+$ state (empty hexagon) by releasing an
electron along the DNA.  The electron is emitted to the left or right
of the enzyme with equal probability. Guanine radicals (circumscribed
dots) can absorb electrons and prevent oxidation of nearby adsorbed
enzymes.  (b) A MutY repair enzyme in the $3+$ state absorbs an
electron and is reduced, causing it to desorb.  (c) Lesions also
prevent passage of electrons, either through electron absorption or
reflection. In our analysis, lesions act differently from oxoG
radicals in that they can continuously absorb electrons.}
\label{fig:fig1}
\end{figure}

The problem of how a BER enzyme locates a lesion on DNA is a specific
example of how enzymes find localized targets.  The DNA of
\textit{E. coli} contains about $10^{6}$ base pairs.  If we assume
that BER enzymes find lesions through a pure 1D diffusive ``sliding''
process with diffusion constant $D \text{~base pairs}^2/s$, the search
time is roughly $10^{12}/D$. Estimating $D$ to be $5 \times 10^{6}
\text{~base pairs}^2/s$, the value for a human DNA glycosylase,
\cite{XIE} we obtain a search time of about $2 \times 10^{5}s \approx
2$ days, much longer than even the reproductive period of
\textit{E. coli}. Therefore, it is likely that other mechanisms are
responsible for DNA target location.

In 1970, Riggs \textit{et al.} \cite{Riggs1,Riggs2} measured the
association rate of the LacI repressor protein to its target on DNA to
be about $10^{10} M^{-1} s^{-1}$. This was puzzling because the
theoretical upper limit for the association rate of a LacI enzyme
diffusing in 3D is predicted (via the Debye-Smoluchowski formula) to
be about 2 orders of magnitude less.  This fundamental biophysical
problem was studied in the seminal work of von Hippel and co-workers
\cite{Berg81,Winter89,vonHippelBerg89,BergvonHippel87} and the
``faster-than-diffusion'' search of targets on DNA has received recent
attention.
\cite{MirnyNature,WunderlichMirny,Klenin,SlutskyMirny,Grosberg,Cherstvy}
Facilitated diffusion is one mechanism
\cite{vonHippelBerg89,SlutskyMirny,WunderlichMirny,HalfordMarko}
proposed to explain the accelerated search. Instead of diffusing
directly to their target, the searching enzymes can spend part of
their time attached to the DNA and perform a 1D random walk along part
of the strand. If the enzyme is able to spend 50\% of its time on the
DNA and 50\% of its time diffusing in 3D, and the diffusion constants
in 1D and 3D are comparable, the association rate is predicted to
increase by as much as 100, \cite{SlutskyMirny} bringing it in line
with the experiments in Riggs \textit{et al.} \cite{Riggs1,Riggs2}
However, other authors have shown that (i) typical enzymes are highly
associated with DNA, spending over 99.999\% of their time on the
strand \cite{WunderlichMirny} and (ii) the diffusion constant in 1D
can be 1000 times smaller than in 3D,\cite{Wang} resulting in a
negligible reduction of the search time. Hence, facilitated diffusion
in its basic form is not adequate to explain the fast reaction rates
observed. Extensions to the facilitated diffusion theory can
incorporate finite enzyme concentrations, \cite{Cherstvy} ``antenna''
effects resulting from the conformation of the DNA, \cite{Grosberg}
fast intersegment transfers of the protein, \cite{SlutskyMirny}
specific/non-specific protein-DNA interactions, \cite{SlutskyMirny}
and directed DNA sliding.\cite{Loverdo}

Although BER enzymes may colocalize to lesions by exploiting the
facilitated diffusion mechanisms cited above, other mechanisms are
likely required for efficient and timely recruitment to lesions.  A
charge-transport (CT) mechanism has been recently proposed as a
possible basis for efficient scanning by MutY, a type of DNA
glycosylase.\cite{BARTON,BARTON2} MutY is known to contain an
iron-sulfur cluster which plays a key role in the CT mechanism. The
cluster can take one of two forms: [4Fe-4S]$^{2+}$ and
[4Fe-4S]$^{3+}$. When MutY is in solution, the cluster is in the $2+$
state and is resistant to oxidation. However, upon binding to DNA, the
cluster potential is shifted, making the $3+$ state more accessible.
The result is that after binding, MutY-[4Fe-4S]$^{2+}$ is easily
oxidized and releases an electron along the DNA, as shown in Fig
\ref{fig:fig1}(a).  It should be noted that the $3+$ state of MutY has
a binding affinity that is about 4 orders of magnitude larger than
that of the $2+$ state.  \cite{Boal} Therefore MutY-[4Fe-4S]$^{2+}$
spends most of its time in solution whereas MutY-[4Fe-4S]$^{3+}$
exists primarily adsorbed onto DNA.

Although controversial about 15 years ago, long range electron
transport in DNA is now a well accepted
phenomenon.\cite{Giese,Schuster} Experiments indicate that charge
transport can occur over $40 \mbox{\AA}$ (about 12 base pairs) in less
than a nanosecond \cite{BARTON3,BARTON4} and the influence of DNA
strand crossovers on CT is generally small. \cite{Giese} Although
electron dynamics along DNA is in general very complicated, some
aspects of the process are now understood.  For example, both guanine
and adenine can act as carriers of positive charge; in analogy with
semiconductors, oxidized DNA can transport charge via the transfer of
holes from base to base. \looseness=-1
%

Quantifying how BER enzymes adsorb to DNA and how they are recruited
to lesions has so far been restricted to simple scaling
arguments.\cite{ERIKSEN} In this paper, in order to explore the
implications of DNA target selection solely by CT, we assume that
adsorbed MutY BER enzymes do not slide along the DNA.  However, upon
first attachment to DNA, the enzyme will emit an electron that
propagates along the strand in a random direction and its cluster will
go from the [4Fe-4S]$^{2+}$ to the [4Fe-4S]$^{3+}$ state.  Should this
electron become absorbed by another MutY-[4Fe-4S]$^{3+}$ enzyme
further along the DNA, the $3+$ form is reduced and desorbs (Fig
\ref{fig:fig1}(b)).  If the electron back-scatters and returns to the
original MutY, it self-desorbs.  Although the model proposed in this
paper is intended to specifically describe the colocalization and
redistribution of MutY through the redox reaction of its iron-sulfur
cluster, many BER enzymes, in fact, contain such a cluster,
\textit{e.g.} endonuclease III. Therefore, we think that our model may
be more general and could also describe the binding kinetics of other
enzymes.

Since unbiased stochastic motion in 1D always leads to return of the
electron, \cite{REDNER} in the absence of any other electron absorbers
on the DNA, a MutY BER enzyme that is deposited will eventually
self-desorb with probability 1. However, BER enzymes can be recruited
to DNA by preexisting electron absorbers. These are typically guanine
radicals (``oxoG'') and other lesions, indicated in
Fig. \ref{fig:fig1}(c) by circumscribed dots and filled diamonds,
respectively. It has been suggested that oxoG plays an important role
in the seeding of MutY onto DNA.\cite{BARTON} The oxoG radicals, like
adsorbed enzymes, are able to absorb electrons, preventing them from
returning and desorbing BER enzymes that originally released them.
Therefore, the oxoG radical in Fig. \ref{fig:fig1}(a) can absorb one
left-moving electron and prevent it from back-scattering and desorbing
the right-most enzyme. Upon reduction, oxoG radicals convert to normal
guanine bases, no longer absorb electrons, and no longer take part in
the CT mechanism.

Other lesions do not simply annihilate by absorbing electrons; rather,
they require the physical presence of BER enzymes to excise them.
These lesions may recruit smaller, more abundant proteins from
solution that permit multiple electron absorption. Another possibility
is that the lesions reflect electrons. Both cases are shown in
Fig. \ref{fig:fig1}(c).
Therefore, our basic model consists of right and left-moving
electrons, guanine radicals, oxidized and reduced forms of BER
enzymes, and lesions on the DNA strand. Newly adsorbed BER enzymes
instantly release electrons (right or left-moving), while oxoG
radicals, lesions, and oxidized BER enzymes absorb electrons and
prevent their passage. 


In this paper, we model the adsorption, desorption and redistribution
of repair enzymes using the redox mechanism shown in
Fig. \ref{fig:fig1}.  We first derive some exact results in the
absence of any lesions; in particular, enzyme adsorption probabilities
and the time taken for returning electrons to induce enzyme
desorption.  These results enable us to define rules for Monte-Carlo
simulations of the dynamics of multiple enzymes.
%
For electron absorbing lesions, simulations show that if enzymes are
deposited onto a DNA at a rate that is slow compared to the electron
dynamics, the distance between a lesion and the closest enzyme scales
as $O(n^{-2/3})$ for large $n$, while total number of enzymes adsorbed
between two lesions scales as $O(n^{1/3})$. However, because of the CT
mechanism, this accumulation is not uniform along the DNA and
the maximum enzyme density always occurs at lesions. Hence \emph{for
electron-absorbing lesions}, the CT mechanism concentrates enzymes to
damaged bases in DNA, consistent with the qualitative predictions in
Yavin, \textit{et al.} \cite{BARTON} and Boon, \textit{et al.}
\cite{BARTON2}

The outline of this paper is as follows.  In the next section, we
develop a model for the electron dynamics based on the stochastic
Broadwell model.  \cite{BICOUT,BROADWELL1,BROADWELL2,SMEREKA} Pairs of
guanine radicals, BER enzymes or lesions define the boundary of a
segment (a ``gap'') over which an electron can propagate.  Section III
contains our results. In Section \ref{subsec:stats}, we we derive
enzyme sticking probabilities and the time taken for returning
electrons to desorb the enzymes that originally emitted them.  In
particular, we derive the MutY desorption rate in terms of the
electron scattering (flip rate) and the electron speed.
In Section \ref{subsec:colocal}, we perform implicit-electron
Monte-Carlo simulations to study the redistribution and accumulation
of enzymes between two fixed lesions on the DNA.  Finally, in Section
IV, we discuss facilitated recruitment of enzymes to lesions in the
context of the CT hypothesis, as well as the biological advantages and
disadvantages of the proposed CT mechanism.

%

\section{Stochastic Charge Transport Model}
\subsection{One-sided Broadwell problem}
\label{subsec:reducedprob}

\begin{table}[htbp]
\begin{tabular}{|c|c|c|}
\hline
Symbol & Definition & Units \\
\hline \hline
$P_+$ & \parbox[c]{2in}{ \vgap Probability density of 
rightward electron \vgap } & 1/L \\
$P_-$ & \parbox[c]{2in}{ \vgap Probability density of 
leftward electron \vgap } & 1/L \\
$X$ & Position along DNA & L \\
$X_0$ & Position of electron release & L \\
$T$ & Time & T \\
$\rho$ & Density of oxoG guanine radicals on DNA & 1/L \\
$L$ & Distance between two oxoGs/enzymes & L \\
$F$ & electron flip rate  & 1/T \\
$V$ & Electron speed & L/T \\
$M$ & Electron decay rate & 1/T \\
$k_{\textrm{on}}$ & Deposition rate of enzymes & 1/(L $\cdot$ T) \\
\hline
\end{tabular}
\caption{Table of dimensional variables and parameters. 
The analysis performed assumes $M=0$.
L represents length and T represents time.}
\label{tab:defn1}
\vspace{0.3cm}
\begin{tabular}{|c|c|c|}
\hline
Symbol & Math defn. & Descriptive definition\\
\hline \hline
$Q_{\pm}$ & $P_{\pm}/\rho$ & \parbox[c]{2in}{ \vgap Rightward/
Leftward electron probability density \vgap} \\
$x$ & $\rho X$ & Coordinate along DNA\\
$x_0$ & $\rho X_0$ & Position of electron release \\
$t$ & $\rho V T$ & Time \\
$\ell$ & $\rho L$ & \parbox[c]{2in}{ \vgap ``Gap size'' : 
distance between two oxoGs/enzymes/lesions \vgap} \\
$f$ & $F/\rho V$ & Electron flip rate \\
$\mu$ & $M/\rho V$ & Electron decay rate \\
$\xi$ & - & Position of enzyme adsorption \\
$d_1, d_2$ & - & \parbox[c]{2in}{ \vgap Enzyme-lesion/enzyme-enzyme 
distance (see Fig. \ref{fig:d1d2}) \vgap} \\
\hline
\end{tabular}
\caption{Definitions of dimensionless symbols in terms of the
dimensional quantities in Table \ref{tab:defn1}.}
\label{tab:defn2}
\end{table}

\begin{figure}[ht]
\begin{center}
\includegraphics[width=3.4in]{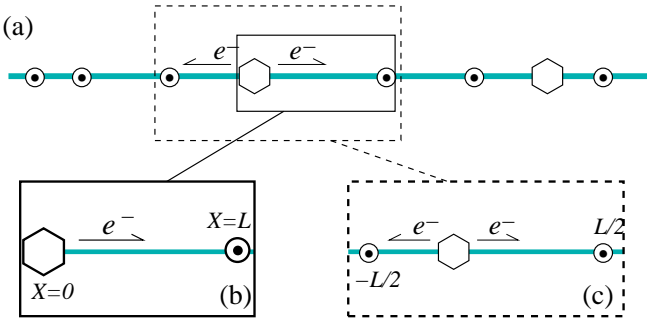}
\end{center} 
\caption{(a) A repair enzyme (hexagon) adsorbs onto a DNA which is
initially populated by guanine radicals (circled dots) with a density
$\rho$.  An electron is emitted to the left or right with equal
probability. The emitted electron has flip rate $F$, speed $V$, and
decay rate $M$.
(b) The one-sided Broadwell problem. An electron
is emitted from $X=0$ with probability $1$ toward a guanine radical at $X=L$.
(c) The two-sided Broadwell problem. An enzyme is deposited between
two guanine radicals which are a distance $L$ apart.  Immediately
after landing inside this segment, an electron is emitted to the left
or right with equal probability.  }
\label{fig:fig2}
\end{figure}

In analogy with Bicout's analysis for the unrelated problem of
microtubule growth dynamics,\cite{BICOUT} we now present 
similar equations for the dynamics of electrons associated with repair
enzymes.
Consider Fig. \ref{fig:fig2}(a): oxoG guanine radicals with density
$\rho$ are distributed randomly along an infinite strand of DNA. A
single repair enzyme initially attaches to the DNA at a random
position, in between two electron absorbing oxoGs. The enzyme
immediately emits an electron along the DNA to the left or right with
equal probability.  The electron can only move with speed $V$, in the
positive or negative $X-$directions, executing random flips between
the two directions with rates $F$.
Furthermore, emitted electrons can be annihilated with rate $M$
through nonspecific interactions with random electron absorbers
diffusing in the bulk.

In general, two steps are required for a MutY enzyme to bind to DNA.
First, when MutY-[4Fe-4S]$^{2+}$ is in contact with the DNA, it has to
undergo oxidation by releasing an electron.  The oxidized form of the
enzyme binds more strongly to DNA.  Second, the released electron must
be absorbed by some particle other than the enzyme (an oxoG, an
already adsorbed MutY or a lesion) to prevent it from returning and
reducing the enzyme.  This allows the enzyme binding to become
``permanent''.
Therefore the net binding probability depends on (i) the probability
of electron release by MutY-[4Fe-4S]$^{2+}$ (when in contact with the
DNA) and (ii) how far neighboring electron absorbers are from the
adsorbed MutY. In this paper, we assume that when enzymes adsorb onto
the DNA, they \emph{always} oxidize, releasing an electron with
probability 1; in Fig. \ref{fig:fig2}(b), the electron is released to
the right with probability 1 and in Fig. \ref{fig:fig2}(c), the
electron is released to the left or right with probability 1/2.  In
principle, an enzyme can attach to and then immediately detach from
the DNA without releasing its electron, but assuming the electron
release rate is large, we neglect this process. The adsorption
probabilities we derive later in this section will depend only on the gap
size $L$ and the parameters for electron motion.
%

Finally, we assume that the DNA is immersed in an infinite reservoir
of enzymes which is kept at a fixed chemical potential.  The rate of
deposition of enzymes onto the DNA is assumed to be constant. A
deposited enzyme can either adsorb by having its released electron
captured by neighboring electron absorbers or it can desorb due to its
electron returning.

To build our full solution, we first derive exact analytical
expressions for the ``one-sided'' problem shown in
Fig. \ref{fig:fig2}(b) which consists of an enzyme at $X=0$ and a
guanine radical at $X=L$.  At time $T=0$, an electron is emitted from
a position $X_0 > 0$ (subsequently, we will take the limit $X_0 \to
0$) with speed $V$ in the positive $X$-direction. For the
one-sided problem, the electron is emitted only to the right.  The
probability that the electron is at a position between $X$ and $X+\dd
X$ at time $T$, and moving to the right with velocity $V$ is denoted
$P_{+}(X,T)$. Similarly, $P_{-}(X,T)$ denotes the probability density
of an electron moving with speed $V$ in the negative
$X$-direction.  The electron can flip directions by scattering from
inhomogeneities and thermally excited conformational variations along
the DNA.\cite{MDO0,MDO1} We model this flipping process as a spatially
homogeneous process occurring with constant rate $F$,
independent of any structure along the DNA such as base pair sequence.

The evolution equations for the probability densities
$P_{\pm}(X,T)$ are

\begin{equation}
%
\begin{array}{rl}
\displaystyle \pt{P_+}{T} &= \displaystyle -V \pt{P_+}{X} - F P_+ + F P_- - M P_+,\\[13pt]
\displaystyle \pt{P_-}{T} &= \displaystyle ~V \pt{P_-}{X} + F P_+ - F P_- - M P_-
\end{array}
\label{EQNP}
\end{equation}
where $0 \leq X \leq L$.
Eqs. (\ref{EQNP}) describe the probability density of electrons being
advected to the right and left.  The flipping of the electron motion
is represented through $F$ and couples the equations for $P_+$ and
$P_-$.  Furthermore, the densities decay in time with an annihilation
rate $M$. Electrons can be annihilated by being absorbed by other
proteins (besides BER enzymes) in solution. If these proteins adsorb
onto the DNA, absorb an electron and desorb back into solution, an
electron is permanently removed from the DNA.

The boundary conditions and initial conditions are  
\begin{eqnarray}
P_{+}(0,T) = P_{-}(L,T) &=& 0, \label{eqn:bc}\\
P_{+}(X,0) &=& \delta(X-X_{0}), \label{eqn:ic1}\\
P_{-}(X,0) &=& 0. \label{eqn:ic2}
\end{eqnarray}
The boundary conditions (\ref{eqn:bc}) arise because the enzyme at
$X=0$ and the oxoG at $X=L$ (see Fig. \ref{fig:fig2}(b)) are both
perfect electron absorbers.  When $X_0 \to 0$, the initial condition
(\ref{eqn:ic1}) reflects the fact that an electron is released to the
right from the enzyme at $X=0$. Initially, there are no leftward
traveling electrons in Fig. \ref{fig:fig2}(b), justifying
Eq. (\ref{eqn:ic2}). All variable and parameters are listed in 
Tables \ref{tab:defn1}.

We now define dimensionless independent variables through the guanine
radical density $\rho$ and the rightward electron travel time $1/(\rho
V)$:
\begin{equation}
\begin{array}{cc}
x = \rho X, & t = \rho V T, \label{eqn:nondim}
\end{array}
\end{equation}
so that Eqs. (\ref{EQNP}) can be written in the form
\begin{equation}
\pt{\mathbf{Q}}{t}  = \mathbf{L} \mathbf{Q},~~~~
{\bf Q} = \left(\begin{array}{c} Q_{+}(x,t) \\[13pt]  
Q_{-}(x,t) \end{array}\right),
\label{eqn:master} 
\end{equation}
where $Q_{\pm} = P_{\pm}/\rho$ and
\begin{equation}
\bf{L} = \left[ \begin{array}{cc}
\displaystyle -\pt{}{x} - f - \mu   &   f \\
f   &   \displaystyle  \pt{}{x} - f - \mu
\end{array} \right], \label{eqn:boldL}
\end{equation}
and $0 \leq x \leq \ell \equiv \rho L$.
In Eq. (\ref{eqn:boldL}),
\begin{equation}
\begin{array}{cc}
\displaystyle f = \frac{F}{\rho V}, \quad & 
\displaystyle \mu = \frac{M}{\rho V},
\end{array}
\end{equation}
is the dimensionless flipping rate and electron
decay rate.  The boundary and initial conditions (\ref{eqn:bc}),
(\ref{eqn:ic1}), (\ref{eqn:ic2}) become
\begin{equation}
\begin{array}{rl}
Q_+(0,t) = Q_-(\ell,t) &= 0, \\
Q_+(x,0) &= \delta(x-x_0), \\
Q_-(x,0) &= 0,
\end{array}
\label{eqn:BC1}
\end{equation}
where $x_0 = \rho X_0$. In the physical problem, an electron is
released from the enzyme as soon as it initially attaches to the
DNA. Therefore, we solve Eqs. (\ref{eqn:master}) with (\ref{eqn:boldL}) and
(\ref{eqn:BC1}) taking the limit $x_0 \rightarrow 0$
(for details, see Appendix \ref{subsec:broadsoln}). 
The dimensionless variables are tabulated and defined in 
Table  \ref{tab:defn2}. Henceforth all of our results and
analyses will be presented for $\mu=0$.

The probability of the enzyme in Fig.~\ref{fig:fig2}(b) self-desorbing
before time $t$ is given by $\int_0^{t} Q_{-}(0,t')\dd t'$ where $Q_-$
can be found by taking the inverse Laplace Transform of Eq. (\ref{A4})
in Appendix \ref{subsec:broadsoln}.  Therefore, the enzyme desorption
and sticking probabilities for the one-sided problem are
\begin{eqnarray}
\frac{f \ell}{1 + f \ell} \quad \mbox{and} \quad 1-\frac{f \ell}{1 + f \ell} 
= \frac{1}{1 + f \ell}, \label{eqn:gamblers}
\end{eqnarray}
respectively. 

\subsection{Two-sided Broadwell problem}

\begin{figure}[htbp]
\includegraphics[width=3.7in]{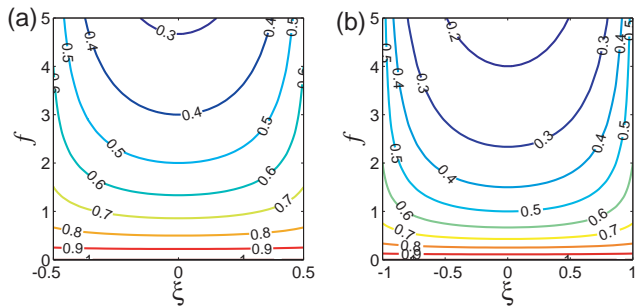}
\caption{
Dependence of enzyme sticking probability,
$\Pi_{\textrm{adsorb}}$ (see Eq. (\ref{eqn:Qadsorb})), 
on dimensionless flip rate $f$
and landing position $-\ell/2 < \xi < \ell/2$,
for the deposition of a single enzyme into a gap (the segment
of DNA between two guanine radicals) of size $\ell$.
(a) $\ell = 1$ with radicals located at $\pm 0.5$ (b) $\ell = 2$ with radicals located at 
$\pm 1$.}
\label{fig:phase_plane}
\end{figure}

Now consider the two-sided problem depicted in Fig. \ref{fig:fig2}(c).
A repair enzyme lands at position $\xi$ between two oxoG guanine
radicals that are a distance $\ell$ apart. The solution to the full
problem can be found by splitting it into two subproblems and using
our results from Section \ref{subsec:reducedprob}. Instead of
solving for the densities on $[0,\ell]$, we can solve for $Q_{\pm}$ separately on
$[\xi,\ell/2]$ (with the enzyme initially deposited at $\xi$ and the
guanine radical at $\ell/2$), on $[\xi,-\ell/2]$ (with the enzyme
at $\xi$ and the guanine radical at $-\ell/2$) and combine the results.
%
%
The enzyme  desorption and adsorption probabilities (\ref{eqn:gamblers})
extend straightforwardly:
\begin{equation}
\begin{array}{l}
\Pi_{\textrm{desorb}}(\xi,\ell) = \displaystyle \frac{1}{2} 
\left[\frac{f(\frac{\ell}{2}-\xi)}{1 + f({\frac{\ell}{2}}-\xi)} +
\frac{f(\frac{\ell}{2}+\xi)}{1+f({\frac{\ell}{2}}+\xi)} \right], \\ [15pt]
\displaystyle \Pi_{\textrm{adsorb}}(\xi,\ell) = \displaystyle
\frac{1}{2} \left[\frac{1}{1 + f({\frac{\ell}{2}}-\xi)} +
\frac{1}{1+f({\frac{\ell}{2}}+\xi)} \right]. 
\label{eqn:Qadsorb}
\end{array}
\end{equation}
A plot of the sticking probability $\Pi_{\textrm{adsorb}}$ for
different values of $f$ and for two different gap sizes is shown in
Fig. \ref{fig:phase_plane}.  For a fixed gap size, and sufficiently
large $f$ (corresponding to a diffusive electron motion), permanent
BER enzyme adsorption is less likely to occur near the center of the
gap because absorption of the electron by guanine radicals is less
likely to occur.  The permanent adsorption or sticking probability is
more uniform when $f$ is small (corresponding to a ballistic electron
motion): whether the oxoG radical is close or far away from the enzyme
makes little difference to the adsorption probability. Finally, for
fixed $f$, increasing the gap size decreases the adsorption
probability because guanine annihilation by the electron is less
likely to occur. The diffusive and ballistic behaviors of the Broadwell
model are derived in Appendix \ref{subsec:limit}.

\begin{table}
\begin{tabular}{|c|c|c|}
\hline
Symbol & Descriptive definition & See Eq. \\
\hline \hline
$\Pi_{\textrm{adsorb}}(\xi,\ell)$ & Enzyme adsorption probability &  (\ref{eqn:Qadsorb})\\
$\Pi_{\textrm{desorb}}(\xi,\ell)$ & Enzyme desorption probability & (\ref{eqn:Qadsorb})\\
$\bar{\Pi}_{\textrm{adsorb}}(\ell)$ & \parbox[c]{2in}{ \vgap Enzyme adsorption prob. 
averaged over landing position $\xi$ \vgap} & (\ref{eqn:invtanh}) \\
$\bar{\Pi}_{\textrm{desorb}}(\ell)$ & \parbox[c]{2in}{ \vgap Enzyme desorption prob. 
averaged over landing position $\xi$ \vgap} & (\ref{eqn:one_minus_invtanh}) \\
$\langle \bar{\Pi}_{\textrm{adsorb}} \rangle $ & \parbox[c]{2in}{ \vgap Enzyme adsorption 
prob. averaged over
landing posn. $\xi$ and gap size $\ell$ \vgap} & (\ref{eqn:angleQ},\ref{eqn:Ei})\\
$t_r$ & \parbox[c]{2in}{\vgap Random variable for conditional 
return time of electron \vgap} & (\ref{eqn:ratio}) \\
$\tau_r(\xi,\ell)$ & \parbox[c]{2in}{\vgap Mean conditional return time (MCRT) of an electron \vgap} & (\ref{eqn:tau})\\ 
$\bar{\tau}_r(\ell)$ & \parbox[c]{2in}{\vgap MCRT of an 
electron averaged over landing position $\xi$ \vgap} & (\ref{eqn:return_time1})\\
$\langle \bar{\tau}_r \rangle$ & \parbox[c]{2in}{\vgap MCRT of an electron averaged over
landing posn. $\xi$ and gap size $\ell$ \vgap} & (\ref{eqn:barbarF}) \\
\hline
\end{tabular}
\caption{Derived adsorption/desorption probabilities, electron return times
and related quantities.}
\label{tab:defn3}
\end{table}

\section{Results and Discussion}

\subsection{Statistics of repair enzymes away from lesions}
\label{subsec:stats}

In this section, we present and discuss deposition statistics that are
valid far away from lesions.  First, using Eq. (\ref{eqn:Qadsorb}), we
average over the landing position $\xi$ to calculate mean
sticking/adsorption probabilities of repair enzymes that are deposited
between two guanine radicals that are a distance $\ell$ apart. The
inter-radical distances (``gaps'') in DNA will, in general, be
randomly distributed. Therefore we ensemble-average our results over
the distribution that $\ell$ is expected to obey. Second, we find the
mean return times of electrons, \textit{i.e.} the time taken for a
deposited enzyme to be desorbed by its own electron, providing it
desorbs.  Again, our results are ensemble-averaged over randomly
distributed gap sizes.  The quantities we shall compute and analyze in
this section are listed in Table \ref{tab:defn3}.

All the results presented are for \emph{adiabatic} depositions. A
deposition is adiabatic if the inter-deposition time is much larger
than the time scale of the electron dynamics. In other words, for
every enzyme deposited, its released electron completes its motion
before the deposition of the next enzyme. At any given time, there is
at most one traveling electron on the DNA. For details, see Appendix
\ref{subsec:adiabatic}.

\subsubsection{Repair enzyme sticking probability}
\label{subsec:adsorption_prob}
One quantity of interest is the probability that any given repair
enzyme that lands on the DNA will not be kicked off by its own
electron, and will remain adsorbed. Enzyme sticking relies on
efficient capture of the released electron by neighboring electron
absorbers (guanine radicals and adsorbed enzymes). Intuitively, one
would expect that a greater density of absorbers with smaller gaps
would result in a more efficient capture of enzymes.

\begin{figure}[htb]
\begin{center}
\includegraphics[width=3.0in]{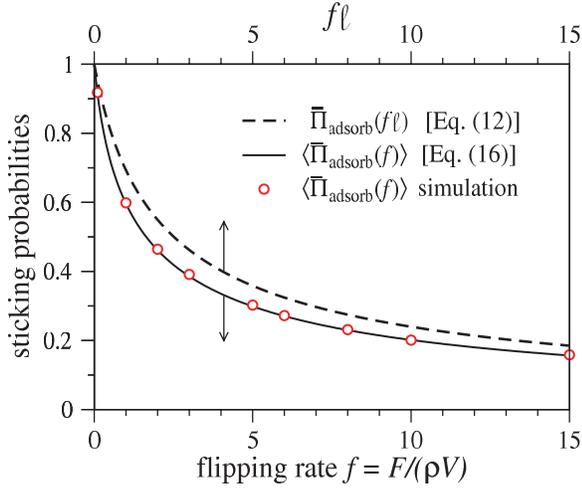}
\end{center} 
\caption{Enzyme sticking probabilities as a function of dimensionless
flip rate $f$.  The simulation data was obtained by performing single
depositions onto a DNA of length 100 ({\it i.e.}, with physical length
$100/\rho$). The fraction of enzymes that remain on the DNA after
performing $10^{5}$ trials was recorded. Increasing the DNA length did
not significantly affect the simulation results.}
\label{fig:Padsorb}
\end{figure}

For a single repair enzyme deposited onto the DNA, landing at a
position $-\ell/2 < \xi < \ell/2$ (see Fig. \ref{fig:fig2}(c)) inside
a gap of length $\ell$, centered about $x=0$, the probability of it
remaining on the DNA is given by $\Pi_{\textrm{adsorb}}$ in
Eq. (\ref{eqn:Qadsorb}).  This quantity can be averaged over all
possible deposition positions $\xi$ within the gap to obtain
\begin{equation}
\begin{array}{rl}
\bar{\Pi}_{\textrm{adsorb}} &\displaystyle = \frac{1}{2 \ell}\int_{-\ell/2}^{\ell/2} \!\!
\left[\frac{1}{1 + f({\ell\over 2}-\xi)} + \frac{1}{1+f({\ell\over 2}+\xi)} \right]
\dd \xi \\[13pt]
\: &\displaystyle = \frac{2}{f \ell} \tanh^{-1}\left(\frac{f \ell}{2+f \ell}\right).
\label{eqn:invtanh}
\end{array}
\end{equation}
This result is plotted in Fig. \ref{fig:Padsorb} (dashed line).
Eq. (\ref{eqn:invtanh}) gives the sticking probability of a repair
enzyme newly deposited between two electron absorbers separated by
$\ell$, {\it uniformly averaged} over its deposition position within
the gap.

We now average over the gap length distribution to compute the
sticking probability for deposited enzymes that land anywhere along
the entire DNA strand. For an infinite, lesion free DNA, depositing an
enzyme will, in general, change the local guanine and enzyme
distribution. Hence, the sticking probabilities will also change with
each successive deposition, making the calculation difficult in the
context of the Broadwell model.  However it is possible to calculate
the sticking probability for a given gap distribution. In special
cases where this distribution is known or simple to calculate, we can
compute the efficiency of enzyme recruitment onto the DNA.

Consider the case of a DNA with a discrete distribution of gaps
$\ell_1,\ell_2,\ell_3,...$. Suppose that on the DNA, a fraction
$\phi_j$ of the gaps have size $\ell_j$.  Now consider many
realizations of a single enzyme deposited onto this DNA.  The fraction
of enzymes that lands in gaps of size $\ell_j$ is $\phi_j \ell_j /
\sum_{j=1}^{\infty} \phi_j \ell_j$ and the fraction of these that
stays adsorbed, using Eq. (\ref{eqn:invtanh}), is
\begin{equation}
\frac{2}{f} \frac{\phi_j \tanh^{-1} \bfrac{f \ell_j}{2+f \ell_j}}{\sum_{j=1}^{\infty} \phi_j \ell_j}.
\end{equation}
The fraction of enzymes that stays adsorbed (in any gap) is obtained
by summing over $j$. In the continuum limit, $\ell_j \to \ell$,
$\phi_j \to \phi(\ell) \dd \ell$, where $\ell$ is the continuous gap
length and $\phi(\ell)$ is the probability distribution function (PDF)
for $\ell$. We obtain
\begin{equation}
\langle\bar{\Pi}_{\textrm{adsorb}}\rangle = \frac{2}{f \langle \ell \rangle} 
\int_0^{\infty} \!\!\! \phi(\ell) \tanh^{-1}
\!\bfrac{f \ell}{2+f \ell} \dd \ell. \label{eqn:angleQ}
\end{equation}
Note that $\langle\bar{\Pi}_{\textrm{adsorb}}\rangle \neq
\int_{0}^{\infty}\phi(\ell)\bar{\Pi}_{\textrm{adsorb}}(\ell)$, the
result that one might expect by naively averaging
Eq. (\ref{eqn:invtanh}) over the gap distribution.

Since $\phi(\ell)$ depends on the number of enzymes deposited, it is
time dependent. In principle, one could calculate how $\phi(\ell)$
changes as enzymes are adiabatically deposited.  The corresponding
evolution of the sticking probability is then given by
Eq. (\ref{eqn:angleQ}). One possible way of finding how $\phi(\ell)$
evolves is to use a mean field theory for the particle distributions,
but we leave this as the subject of a future investigation.

In the special case where one enzyme is deposited onto a DNA that only
has guanine radicals, we can calculate $\phi(\ell)$ and hence
$\langle\bar{\Pi}_{\textrm{adsorb}}\rangle$ explicitly. If the guanine
radicals have a number density $\rho$, then the gap lengths, on
average, are $1/\rho$, which corresponds to a unit dimensionless gap
size (see Eq. (\ref{eqn:nondim})).  Hence $\langle \ell \rangle = 1$
and the dimensionless gap sizes, $Y$, are exponentially distributed
(see Appendix \ref{subsec:gap}) according to 
\begin{equation}
\text{Prob}(\ell < \text{gap size} < \ell + \dd \ell) = e^{-\ell} \dd \ell, \label{eqn:expgap}
\end{equation}
so we set $\phi(\ell) = e^{-\ell}$. Substituting this 
result into Eq. (\ref{eqn:angleQ}), we obtain
\begin{equation}
\langle\bar{\Pi}_{\textrm{adsorb}}\rangle = \frac{e^{1/f} \mbox{Ei}(1/f)}{f},  
\label{eqn:Ei}
\end{equation}
where $\mbox{Ei}(x) = \int_x^{\infty} \frac{e^{-t}}{t}\dd t$ is the
exponential integral.  This analytic result is plotted in
Fig. \ref{fig:Padsorb} (solid line) and is confirmed by Monte-Carlo
simulations (circles). The sticking probability increases when either
the electron-absorber density $\rho$ increases, the electron velocity
$V$ increases or the flip rate $F$ decreases.

Equation (\ref{eqn:Ei}) is valid only when the number of enzymes that
have stuck is much less than the initial number of oxoG radicals.  In
this limit, the distribution of gap lengths will remain approximately
exponential.  For the human genome of $\sim 10^9$ base pairs, there
are approximately $10^4$ oxoGs present at any given
time.\cite{Helbock} In this case, we expect that Eq. (\ref{eqn:Ei})
should be fairly accurate for about the first dozen depositions.

Note that $\bar{\Pi}_{\textrm{adsorb}}$ (Eq.~(\ref{eqn:invtanh})) with
$\ell = 1$ gives the enzyme sticking probability inside an
inter-radical gap of unit length, whereas $\langle
\bar{\Pi}_{\textrm{adsorb}} \rangle$ (Eq. (\ref{eqn:Ei})) gives the
enzyme sticking probability averaged over exponentially distributed
inter-radical gaps lengths, but with unit mean.  Intuitively, one
would expect the boundaries defining the smaller gaps to be more
efficient at sequestering electrons than those associated with larger
gaps. However, the enhanced electron trapping by smaller gaps, leading
to otherwise increased sticking probabilities is compensated by a
higher deposition flux into larger gaps (large gaps collect more
enzymes than small gaps). The net result of averaging over
exponentially distributed gap sizes is for the larger gaps to dominate
and lower the overall gap-averaged sticking probability. This is shown
in Fig. \ref{fig:Padsorb} where for all values of $f$, $\langle
\bar{\Pi}_{\textrm{adsorb}} \rangle < \bar{\Pi}_{\textrm{adsorb}}$
when $\ell = 1$. 
\vspace{0.3cm}

\subsubsection{Mean conditional return time of electrons}
 
We now find the mean time that a BER enzyme stays on the DNA after its
initial deposition, conditioned on its own electron returning and
knocking the enzyme off.  This quantity allows us to estimate a rate
of desorption that can be used in more coarse-grained, higher level
descriptions of the CT mechanism.

Consider depositing an enzyme into a gap of size $\ell$ at a position
$\xi$ satisfying $-\ell/2 < \xi < \ell/2$.  The probability that the
electron (``$e^{-}$'') returns in a time $t_{r} < t$, given that it
returns is,

\begin{widetext}
\begin{eqnarray}
\mbox{Prob}(t_{r}  &<& t~|~ e^{-}~\text{returns}) = \frac{\mbox{Prob}(t_r < t)}
{\mbox{Prob}(e^{-}~\text{returns})} \nonumber \\
&=& \frac{\mbox{Prob}(t_{r} < t ~|~ e^{-}~\text{shoots right}) + 
\mbox{Prob}(t_{r} < t ~|~ e^{-}~\text{shoots left}) 
}{\mbox{Prob}(e^{-}~\text{returns} ~|~ e^{-}~\text{shoots right}) +
\mbox{Prob}(e^{-}~\text{returns} ~|~ e^{-}~\text{shoots left})} \nonumber \\
&=& \frac{\frac{1}{2} \int_0^{t} Q_-(x=0,t';0,\ell/2-\xi)\dd t' + \frac{1}{2} \int_0^{t}
Q_-(x=0,t';0,\ell/2+\xi)\dd t'}
{\frac{1}{2} \int_0^{\infty} Q_-(x=0,t';0,\ell/2-\xi)\dd t' + 
\frac{1}{2} \int_0^{\infty} Q_-(x=0,t';0,\ell/2+\xi)\dd t'}. \label{eqn:ratio}
\end{eqnarray}
\end{widetext}
In Eq. (\ref{eqn:ratio}), $Q_-(x,t;x_0,\ell)$ is the leftward electron
density at position $0<x<\ell$ at time $t$ given that the electron was
released from $x=x_0$ at $t=0$ (see Fig. \ref{fig:fig1}(b) for the
$x_0=0$ case).  This density comes from solving
Eqs.~(\ref{eqn:master}) and (\ref{eqn:boldL}) along with the
conditions (\ref{eqn:BC1}).  

\begin{figure}[htbp]
\begin{center}
\includegraphics[width=3.3in]{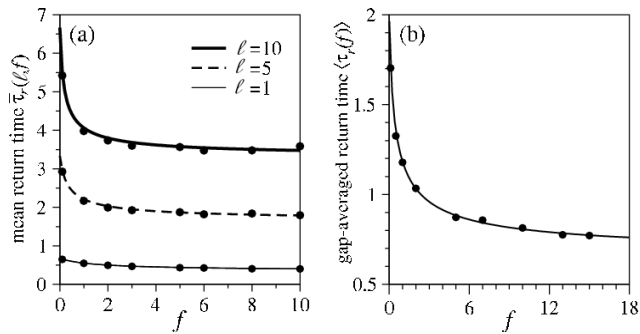}
\end{center} 
\caption{
(a) For adiabatic enzyme depositions into a gap of size $\ell$, 
$\bar{\tau}_r$, the mean conditional return time of
an electron, averaged over the enzyme landing position $\xi$,
is recorded for different gap sizes and 
dimensionless flip rates $f$. The symbols
represent data from Monte Carlo simulations and the solid line
represents the analytic expression from Eq. (\ref{eqn:return_time1}).
(b) For random, uniform, adiabatic enzyme depositions onto a DNA with
randomly and uniformly distributed guanine radicals,
$\langle\bar{\tau}_{r}\rangle$ as predicted by
Eq. (\ref{eqn:barbarF}) is plotted as a function of the dimensionless
flip rate $f$.}
\label{fig:return_time}
\end{figure}

The mean conditional electron return time $\tau_{r}$ can then be
computed from
\begin{equation}
\tau_{r}(\xi; \ell, f) = \int_{0}^{\infty}\!\! t 
{\partial \over \partial t}~ \mbox{Prob}(t_{r}< t \vert e^{-}~\mbox{returns}) \,\dd t.
\label{eqn:tau}
\end{equation}
%
Using Eq. (\ref{eqn:ratio}), $\tau_{r}(\xi;\ell,f)$ in
Eq. (\ref{eqn:tau}) can be found in terms of the Laplace-transformed
density $\tilde{Q}_{\pm}(x,s)$ which is given in Eq.  (\ref{A4}) of
Appendix \ref{subsec:broadsoln}. Upon averaging $\tau_{r}(\xi;\ell,f)$
over the initial landing positions $\xi$, we obtain
\begin{equation}
\begin{array}{l}
\displaystyle \bar{\tau}_{r}(\ell,f) = \frac{2}{3f}
\frac{3+f \ell}{\sqrt{f\ell(2+f \ell)}} \tanh^{-1} \bfrac{\sqrt{f \ell}}{\sqrt{2+f \ell}}
\\[13pt] \:\hspace{2.5cm} \displaystyle +\frac{1}{3f}[f \ell -1 -
\frac{2}{f\ell}\log(1+f\ell)].
\label{eqn:return_time1}
\end{array}
\end{equation}
We plot $\bar{\tau}_{r}(\ell,f)$, and validate
Eq. (\ref{eqn:return_time1}) using MC simulations in
Fig. \ref{fig:return_time}(a).
Finally, we further ensemble-average $\bar{\tau}_{r}$ over gap lengths
$\ell$. Consider many realizations of the deposition of a single
enzyme onto an infinite DNA with oxoGs whose gaps are exponentially
distributed.  The average time that the enzyme stays adsorbed, given
that its electron eventually returns to knock it off, is $\langle
\bar{\tau}_r \rangle$.  The calculation of $\langle \bar{\tau}_r
\rangle$ is similar to that of
$\langle\bar{\Pi}_{\textrm{adsorb}}\rangle$ described in Section
\ref{subsec:adsorption_prob}, but modified to account for the fact
that the number of enzymes that self-desorb ({\it i.e.}, the number of
return times that are finite) depends on $\ell$.  If an enzyme is
deposited into a gap of size $\ell$, the probability of self-desorbing
after a finite time is given by (see Eq. (\ref{eqn:invtanh}))
\begin{equation}
\bar{\Pi}_{\textrm{desorb}}(\ell, f) = 1 - \frac{2}{f \ell} 
\tanh^{-1}\bfrac{f \ell}{2+f \ell}.
\label{eqn:one_minus_invtanh}
\end{equation}
Therefore, the required expression for $\langle\bar{\tau}_r \rangle$ is
\begin{equation}
\langle \bar{\tau}_{r}(f)\rangle = \frac{
\displaystyle \int_0^{\infty} \bar{\Pi}_{\textrm{desorb}}(\ell,f) 
\bar{\tau}_{r}(\ell,f) \ell e^{-\ell} \dd \ell}{
\displaystyle \int_0^{\infty}  
\bar{\Pi}_{\textrm{desorb}}(\ell,f) \ell e^{-\ell} \dd \ell} 
\label{eqn:barbarF}
\end{equation}

%
In the numerator of Eq. (\ref{eqn:barbarF}), $\bar{\Pi}_{\textrm{desorb}}
(\ell,f) \ell e^{-\ell} \dd \ell$ is the fraction of deposited enzymes that (i) land in
a gap that has a length between $\ell$ and $\ell+\dd\ell$ and (ii)
eventually self-desorb after finite time. In the denominator,
$\int_0^{\infty} \bar{\Pi}_{\textrm{desorb}}(\ell,f) \ell e^{-\ell} \dd\ell$ is the fraction of
deposited enzymes that self-desorb after a finite time. The result
(\ref{eqn:barbarF}) is confirmed by simulation data in
Fig. \ref{fig:return_time}(b).

Equation (\ref{eqn:barbarF}) was derived by considering the
deposition of a single enzyme onto an infinite DNA with exponentially
distributed gap lengths.  However, as is the case with
Eq. (\ref{eqn:Ei}), it is also approximately true for a small number
of depositions onto a finite DNA: providing the number of oxoGs
annihilated is small compared to the total number of oxoGs, the
distribution of gap lengths is still approximately exponential.
Hence, for a given deposition rate of enzymes \textit{per unit length}
onto an infinite DNA, Eq. (\ref{eqn:barbarF}) will hold approximately
for times such that the fraction of oxoGs annihilated is small.

Given a deposition rate of enzymes (per unit length), we can estimate
a desorption rate (per unit length) from Eq. (\ref{eqn:barbarF}).  If
desorption were a Poisson process, then the desorption rate,
$k_{\textrm{off}}$, is found from the inverse of the mean unbinding
time of the repair enzyme.  Although the desorption process in our
model depends on the dynamics of electron charge transport (rendering
it to be non-Poisson), the inverse of the ensemble averaged
conditional return time of an electron $1/\langle \bar{\tau}_{r}
\rangle$, is nonetheless a reasonable definition for the detachment
rate $k_{\textrm{off}}$.  We expect this value of $k_{\textrm{off}}$
to be accurate, as long as the fraction of oxoGs annihilated by repair
enzymes is small. The probabilities and times relevant to electron
dynamics are summarized in Table \ref{tab:defn3}.

\subsection{Colocalization of enzymes to lesions}
\label{subsec:colocal}
We now consider a permanent lesion on the DNA (one
that does not annihilate upon absorption of an electron).  Such a
lesion may be bound to other enzymes and cofactors so that it can act
as a sink for multiple electrons, or it can reflect electrons.  In
this section, we consider lesions that can either absorb or reflect
electrons, as shown in Fig. \ref{fig:fig1}(c). We are primarily
interested in the average number of depositions required for a repair
enzyme to be adsorbed within a certain (small) distance from the
lesion.

For the sequential deposition of many enzymes onto a DNA populated
with guanine radicals and lesions, the evolution of enzyme and guanine
densities is not amenable to exact analytical solution.  
Therefore, our approach will be to track
enzyme-lesion distances and enzyme concentrations on the DNA by
performing Monte-Carlo simulations.

Each simulation consists of a series of adiabatic depositions. 
A deposition is simply the spontaneous appearance of a MutY-[4Fe-4S]$^{3+}$ enzyme at
a randomly chosen position along the DNA. Note that a deposition is an
\emph{attempted} adsorption: it can result either in the enzyme sticking to the DNA,
or desorbing from it. In our simulations, the
number of enzymes on the DNA can grow without bound.
We do not model the bulk dynamics
for MutY-[4Fe-4S]$^{2+}$ enzymes in solution.

\begin{figure}[htbp]
\begin{center}
\includegraphics[width=3.3in]{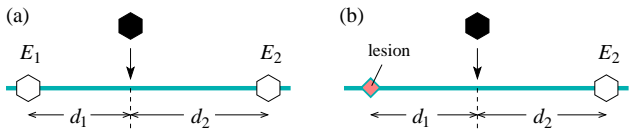}
\end{center}
\caption{(a) Deposition of new enzyme $E$ (solid hexagon) in between
two adsorbed enzymes $E_1$ and $E_2$ (empty hexagons). (b) Deposition
of a new enzyme $E$ between a lesion and an adsorbed enzyme,
$E_2$. Adsorption and desorption probabilities are given in Tables
\ref{table1} and \ref{table2}.}
\label{fig:d1d2}
\end{figure}

In our model, each enzyme that is deposited releases an electron along
the DNA. However, rather than performing time-consuming, explicit
simulations of a Broadwell process, we exploit our analytic results to
implicitly account for the electrons.  The rules for enzyme desorption
and adsorption come from the probabilities $\Pi_{\textrm{desorb}}$ and
$\Pi_{\textrm{adsorb}}$ found in Eqs. (\ref{eqn:Qadsorb}).
Specifically, consider the deposition of an enzyme, $E$, between two
already adsorbed enzymes, $E_1$ and $E_2$ (see
Fig. \ref{fig:d1d2}(a)). Let the distance from $E$ to $E_i$ be $d_i$,
$i=1,2$.  Then the probability of $E$ adsorbing and knocking off $E_i$
is $\frac{1}{2} \frac{1}{1 + f d_i}$ and the probability of $E$
self-desorbing is $\frac{1}{2} ( \frac{ f d_1 }{1 + f d_1 } + \frac{ f
d_2 }{ 1 + f d_2 } )$.  In the case where an enzyme is deposited
between a lesion and an adsorbed enzyme (see Fig. \ref{fig:d1d2}(b)),
the adsorption and desorption probabilities have to be modified. If
$E_1$ is replaced by an electron-reflecting lesion, the probability of
$E$ permanently adsorbing without displacing $E_2$ is zero. The
probability of $E$ adsorbing and knocking off $E_2$ is
$\frac{1}{2}\frac{1}{1+f d_1}$ and the probability of self-desorption
is $\frac{1}{2} + \frac{1}{2} \frac{f d_1}{1+f d_1}$.
If $E_1$ is replaced by an electron-absorbing lesion, the probability
of $E$ permanently adsorbing without displacing $E_2$ is $\frac{1}{2}
\frac{1}{1 + f d_1}$, the probability of $E$ adsorbing and knocking
off $E_2$ is $\frac{1}{2}\frac{1}{1+f d_2}$ and the probability of
self-desorption is $\frac{1}{2}(\frac{f d_1}{1+f d_1} + \frac{f
d_2}{1+f d_2})$.  These probabilities are summarized in Tables
\ref{table1} and \ref{table2}.

\begin{table}[htbp]
\begin{tabular}{|c|c|c|c|}
\hline
Event: & $E$ self-desorbs & \parbox[c]{0.7in}{$E$ adsorbs, $E_1$ desorbs}  
& \parbox[c]{0.7in}{\vgap $E$ adsorbs, $E_2$ desorbs \vgap} \\
\hline
\parbox[c]{0.7in}{\vgap Probability: \vgap} & 
$\frac{1}{2} \left( \frac{f d_1}{1 + f d_1} + 
\frac{f d_2}{1 + f d_2}\right)$ &$ \frac{1}{2} \frac{1}{1+f d_1}$ & $\frac{1}{2}
\frac{1}{1+ f d_2}$ \\
\hline
\end{tabular}
\caption{Adsorption and desorption probabilities in
Fig. \ref{fig:d1d2}(a) when the enzyme $E$ is deposited between
enzymes $E_1$ and $E_2$.}
\label{table1}
%
\vspace{0.3cm}
\begin{tabular}{|c|c|c|c|}
\hline Event: & $E$ self-desorbs & \parbox[c]{0.7in}{\vgap $E$
adsorbs, $E_2$ stays adsorbed \vgap} & \parbox[c]{0.7in}{$E$ adsorbs,
$E_2$ desorbs} \\ \hline
\parbox[c]{0.7in}{\vgap Probability: 
(reflecting lesion) \vgap} & $\frac{1}{2} + \frac{1}{2} \left( \frac{f d_2}
{1 + f d_2} \right)$ & $0$ &$ \frac{1}{2} \frac{1}{1+f d_2}$ \\
\hline
\parbox[c]{0.7in}{\vgap Probability: (absorbing lesion) \vgap} & 
$\frac{1}{2} \left( \frac{f d_1}{1 + f d_1} + \frac{f d_2}{1 + f d_2} \right)$ & $\frac{1}{2}
\frac{1}{1+ f d_1}$ & $\frac{1}{2} \frac{1}{1+f d_2}$\\
\hline
\end{tabular}
\caption{Adsorption and desorption probabilities in
Fig. \ref{fig:d1d2}(b) when the lesion is an electron absorber and
reflector.}
\label{table2}
\end{table}

MC simulations were performed on a periodic domain of size $\Gamma$
containing a single lesion, which is equivalent to a single finite
domain with length $\Gamma$ and lesions at $x=0$ and $x=\Gamma$. We
start our simulations with no adsorbed BER enzyme (MutY), but with a
unit density of guanine radicals (oxoG) whose gaps follow an
exponential distribution (see Eq. (\ref{eqn:expgap})). When a single
enzyme is deposited randomly on $[0,\Gamma]$, the positions of the two
particles (either oxoGs, lesions or already adsorbed enzymes) on
either side are recorded and $d_1$ and $d_2$ are calculated (see
Fig. \ref{fig:d1d2}). Using the probabilities in Tables \ref{table1}
and \ref{table2}, the outcome of this deposition event is determined:
either the newly deposited enzyme adsorbs, or it desorbs due to its
electron returning. Note that if an adsorption occurs, exactly one of
three other events also has to occur: (i) a neighboring enzyme is
reduced and desorbs (ii) a neighboring oxoG is annihilated or (iii) an
electron is absorbed by a neighboring lesion. 

\begin{figure*}
\begin{center}
\includegraphics[width=5.5in]{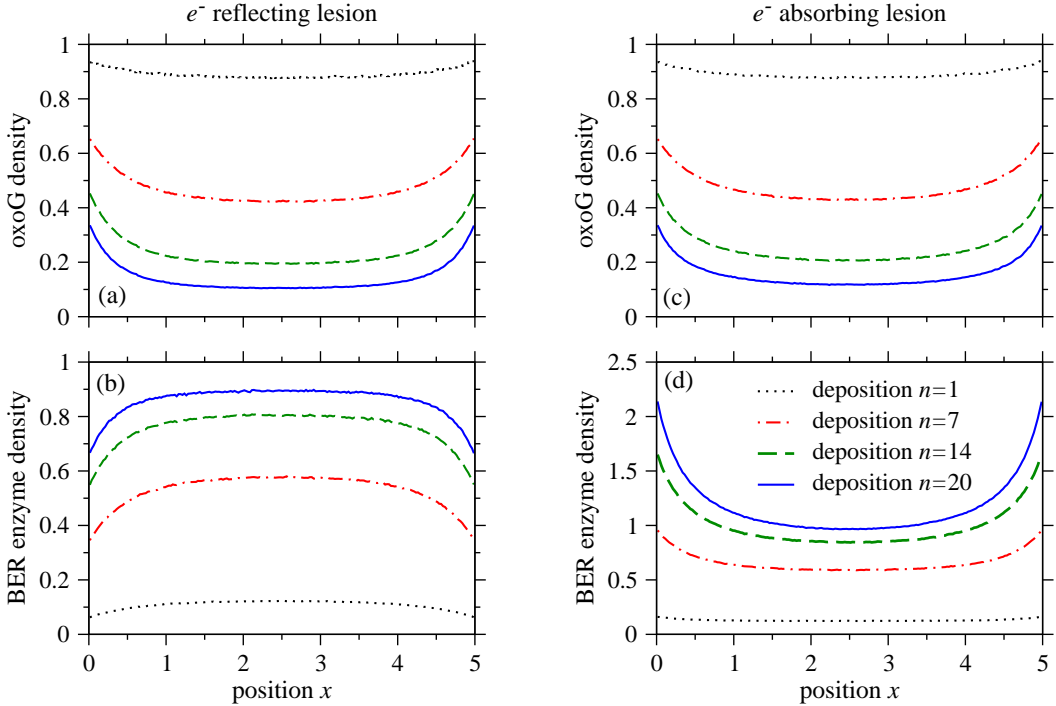} 
\end{center} 
\caption{Implicit-electron Monte-Carlo simulations of the evolution of
mean guanine radical ((a) and (c)) and BER enzyme ((b) and (d))
density profiles after 1 (dotted), 7 (dot-dashed), 14 (dashed), and 20
(solid) enzyme depositions.  (a) and (b) correspond to
electron-reflecting lesions at $x=0$ and $x=\Gamma=5$, and (c) and (d)
are for electron-absorbing lesions.  Results were obtained from
averaging $10^7$ trials and using a flip rate of $f=1$.}
\label{fig:lesion_reflect}
\end{figure*}

Figure \ref{fig:lesion_reflect} shows density profiles obtained from
our MC simulations. In Fig. \ref{fig:lesion_reflect}(a), the depletion
of guanine radicals is greater away from lesions: a guanine radical
that is close to a lesion can, essentially, only be annihilated from
one side.  Near $x=0$, the probability of oxoGs being annihilated from
the left by a rightward-moving electron is very small.  Similarly,
near $x=5$, the probability that oxoGs are annihilated from the right
by leftward-moving electrons is also very small.

Figure \ref{fig:lesion_reflect}(b) shows that electron reflecting
lesions eventually prevent the build up of enzymes near lesions. The
presence of an electron-reflecting lesion increases the local
self-desorption rate. Note that the enzyme self-desorption probability
is always greater in Fig. \ref{fig:d1d2}(b) than it is in
Fig. \ref{fig:d1d2}(a) -- when the lesion is electron reflecting.
Therefore, near a reflecting lesion, the recruitment of enzymes by
guanine radicals has to compete with this increased self-desorption
rate. Although the density near the lesion increases with time, for a
fixed time, its value is always smaller than the bulk value.  Another
way to understand the enzyme depletion is through a particle
conservation argument.  Since the total number of guanine radicals and
BER enzymes is conserved, an increase in oxoG density near the
boundaries must correspond to a decrease in the enzyme density.

%
Figures \ref{fig:lesion_reflect}(c) and \ref{fig:lesion_reflect}(d)
show density profiles near electron-absorbing lesions.  The oxoG
densities in Fig. \ref{fig:lesion_reflect}(c) remain essentially
unchanged from those surrounded by electron-reflecting lesions (Fig.
\ref{fig:lesion_reflect}(a)). As shown in
Figs. \ref{fig:lesion_reflect}(b) and \ref{fig:lesion_reflect}(d), the
BER enzyme density profiles are also similar for a small number of
depositions, away from lesions. On the other hand,
\ref{fig:lesion_reflect}(d) also shows that for larger deposition
numbers, the BER enzyme density near electron-absorbing lesions
increases markedly.

The total number of particles on the DNA strand can be found by
integrating the densities from $x=0$ to $x=\Gamma$.  For example, in
Fig. \ref{fig:lesion_reflect}(b), the solid curve representing the
enzyme density after one attempted deposition takes the value $\sim
0.12$ over most of the domain and decreases slightly near the lesions.
Therefore the (average) number of enzymes that remain adsorbed
after one attempted deposition is approximately $0.12 \times 5 =
0.6$. This is in excellent agreement with the solid curve in
Fig. \ref{fig:Padsorb} and Eq. (\ref{eqn:Ei}) for $f=1$ since $\langle
\bar{\Pi}_{\textrm{adsorb}} \rangle = e \mbox{Ei}(1) = 0.596...$

Figure \ref{fig:lesion_reflect} only shows the densities up to 20
deposition attempts.  When the number of depositions is much greater
than 20, all of the enzyme-seeding guanine radicals are
annihilated. In the absence of any electron absorbers on the DNA,
there can be no net increase in enzyme number, and the enzyme density
in Fig. \ref{fig:lesion_reflect}(b) eventually saturates to unity
everywhere in the domain, identical to the initial oxoG density.  Each
guanine radical is eventually replaced by a BER enzyme, so the
long-time BER enzyme density mimics the initial oxoG density. 

%
In contrast, when the lesions are electron absorbing, there are always
two permanent electron absorbers in the system.  In this case, the
number of enzymes can grow without bound, even when all the oxoGs are
depleted.

\begin{figure}[htb]
\begin{center}
\includegraphics[width=3.4in]{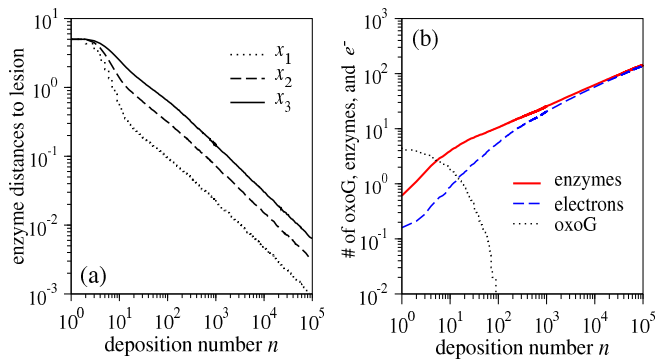} 
\end{center} 
\caption{(a) Convergence of repair enzymes to an electron absorbing
lesion. The distance between the lesion and the closest, second
closest and third closest enzymes (denoted by $x_j$, $j=1,2,3$
respectively) scales as $O(n^{-2/3})$ for $n \gg 1$ where $n$ is the
deposition number. Results were obtained using $f=1$ and by averaging
over 5000 trials. (b) The total number of enzymes and guanine radicals
on the DNA and the number of electrons absorbed by the lesion as a
function of deposition number after averaging over 100 trials. The
enzyme number scales as $O(n^{1/3})$ for $n \gg 1$.  The dimensionless
flip rate was $f=1$, and the domain size was $\Gamma = 5$.  There were
initially 5 guanine radicals present.}
\label{fig:convergence}
\end{figure}

Figure \ref{fig:convergence}(a) shows how enzymes converge to electron
absorbing lesions located at $x=0$ and $x=\Gamma=5$.  At any given
time, we label the $m$ enzymes on the DNA according to their position
$E_i$ so that $0 < E_1 < E_2 < ... < E_m < \Gamma$. Both the number of
enzymes on the DNA, $m$, and their positions, $E_i$, are functions of
$n$, the number of (attempted) depositions that have occurred.
We plot the quantities $x_1 = \min(E_1,\Gamma-E_m)$, $x_2 =
\min(E_2,\Gamma-E_{m-1})$ and $x_3 = \min(E_3,\Gamma-E_{m-2})$ as
functions of deposition number $n$ in Fig. \ref{fig:convergence}(a).
When fewer than 3 enzymes are adsorbed on the DNA, we define $x_i = \Gamma$,
$i=1,2,3$.
From our simulations, we find the scaling
\begin{equation} 
x_i \sim n^{-2/3} \text{~for~} i=1,2,3,  \label{eqn:twothirds}
\end{equation}
in the large $n$ limit. 
%
%
For a BER enzyme to successfully excise a lesion, we assume that it
has to be within a few base pairs of it. We set the physical enzyme-lesion
distance $X_1 \equiv x_1/\rho = 5a$, where $a$ is the width of a base
pair which we take to be $0.34$ nm, and estimate $n$. Approximately 1
in 40,000 guanine bases are guanine radicals, \cite{Helbock} so $\rho
= (160,000 a)^{-1}$, and the number of attempted depositions
required for the closest sticking enzyme to be within 5 base pairs of
the lesion is $n \approx 6 \times 10^6$.  If each deposition takes at
least 0.0005 seconds,
\footnote{
For \textit{E. Coli}, the maximum deposition rate can be obtained by
assuming a nucleoid radius of approximately $b\approx 0.3\mu$m. Upon
assuming a MutY diffusivity of $D\sim 3 \times 10^{-7}$cm$^{2}$/s, the
Debye-Smoluchowski estimate is $k_{\textrm{on}} \sim 4 \pi D b \approx
6 \times 10^{10}$ M$^{-1}$s$^{-1}$ For MutY concentration of $C
\approx \textrm{20 enzymes/fL}$, the average time between depositions
is $(kC)^{-1} \approx 0.0005s$.
}
this amounts to a total (minimum) search time of about 50 minutes.
Although this is a significant reduction compared to the original 1D
sliding search time discussed in the Introduction, it is likely that
MutY locates lesions even more quickly through a combination of the CT
mechanism and facilitated diffusion along the DNA strand.

The solid curve in Fig. \ref{fig:convergence}(b) shows the total
number of enzymes on the DNA as a function of the deposition number
when the lesions at $x=0$ and $x=5$ are electron absorbing.  Upon
depletion of the guanine radicals (shown by the dotted curve dropping
to $< 10^{-2}$), the enzyme number increases as $O(n^{1/3})$.  The
dashed curve in Fig. \ref{fig:convergence}(b) shows the number of
electrons absorbed by the lesion. Initially, this is less than the
enzyme number since enzymes adsorb mainly by oxoG
annihilation. However, as all the radicals are used up, the dashed
curve asymptotes to the curve for the enzyme total, indicating that
the net increase of enzymes on the DNA is due primarily to
lesion-induced colocalization.

Given that the enzyme-lesion distance scales as $O(n^{-2/3})$ for
electron absorbing lesions, one can directly show that the number of
enzymes on the DNA scales as $O(n^{1/3})$ through a simple
argument. If the enzyme-lesion distance is $O(n^{-2/3})$ it takes
$O(n^{2/3})$ attempts before an added enzyme lands closest to the
lesion.  When $n$ is large, the enzyme-lesion distance is small and
the electron released by the newly deposited enzyme will be
absorbed. For every $O(n^{2/3})$ depositions, on average, one
permanent adsorption occurs.  Hence, for every $O(n)$ depositions,
$O(n^{1/3})$ adsorptions occur.

While the convergence of CT enzymes towards lesions scales as $x_i
\sim n^{-2/3}$, the convergence of passive enzymes (those that simply
adsorb onto DNA without emitting electrons) scales as $x_i \sim
n^{-1}$. The faster convergence of passive enzymes \footnote{
While passive enzymes converge more quickly to enzymes when measured
in terms of the deposition number $n$, CT repair enzymes converge more
quickly when measured in terms of the number of adsorptions $m =
O(n^{1/3})$. In fact, the enzyme-lesion distance for repair enzymes
scales as $O(n^{-2/3}) = O(m^{-2})$ compared to $O(m^{-1})$ for
passive enzymes with $m=n$ (every deposition results in an
adsorption).
} is a consequence of linearly increasing the passive enzyme density
on the DNA. The CT mechanism on the other hand, prevents the
recruitment of large numbers of BER enzyme on the DNA at any given
time with the total number scaling as $O(n^{1/3}) \ll O(n)$ for large
$n$. Hence, although BER enzymes only colocalize near lesions (note
the maxima in the enzyme density occur at the lesions in
Fig. \ref{fig:lesion_reflect}(d)), and the CT mechanism suppresses the
wasteful build-up of enzymes in undamaged parts of the DNA.

\section{Summary and Conclusions}

We developed a mathematical model for a proposed charge-transport
mediated mechanism of Base Excision Repair (BER) enzyme colocalization
to DNA lesions.  Enzymes adsorb and desorb through a charge transport
(CT) mechanism \cite{BARTON,BARTON2} which we model using a stochastic
Broadwell process. Our main finding is that the CT mechanism
concentrates repair enzymes at lesions provided the lesions are
electron absorbing.

%

We first calculated enzyme sticking probabilities and self-desorption
rates in the absence of lesions.  
%
%
Our results for an infinite, lesion free DNA,
populated with guanine radicals, are summarized in
Figs. \ref{fig:Padsorb} (which predicts the enzyme sticking probability) and
\ref{fig:return_time} (which predicts the electron's mean conditional return
time).  For the deposition of a single enzyme onto an infinite DNA,
the results are exact; for a given deposition rate per unit length, we
expect the results to hold approximately providing the fraction of
guanine radicals (oxoGs) annihilated is small.
%
%
We also explored how enzymes colocalize to lesions using Monte-Carlo
simulations.  Enzymes were adiabatically deposited onto a circular DNA
with a single lesion. We found that electron-absorbing lesions
colocalize CT enzymes, and while electron-reflecting lesions do not
(Fig. \ref{fig:lesion_reflect}).

Simple faciliated diffusion is often unable to account for the fast
search times observed in certain DNA-protein
reactions. \cite{Riggs1,Riggs2} Cherstvy \textit{et al.}
\cite{Cherstvy} state that under realistic conditions, facilitated
diffusion cannot occur and propose that acceleration is achieved
through the collective behavior of proteins. In the context of target
search by enzymes, the CT mechanism complements facilitated diffusion
models. \cite{SlutskyMirny,vonHippelBerg89,Grosberg} The CT-mediated
mechanism is one such example of collective behavior.  Instead of
basing the enzyme search problem on the time for a single protein to
find its target, the CT mechanism relies on a collective build-up of
enzyme density at the lesion. Hence, issues important in facilitated
diffusion theories, such as the starting point of the enzyme relative
to the lesion and the length of the DNA become irrelevant in the CT
mechanism.

In the case where targets (lesions) are electron absorbing, we find
that the maximum enzyme density always occurs at the permanent lesions
and furthermore that the CT mechanism maintains a low density of
enzymes far from lesions to suppress oxoGs, which are another form
(albeit less permanent) of DNA damage.  In fact, after an initial
transient where all oxoGs are annihilated, the density of enzymes for
most of the DNA will be of the order of the oxoG density, which is
very low (about 1 in 160,000 base pairs).  Subsequent enzyme
depositions will colocalize only near the lesion. Our results show
that although $n \sim 10^6$ (attempted) depositions are required for
the concentration to build up to a sufficient level at the lesions in
order for them to be excised, the number of enzymes actually adsorbed
on the DNA is much less, at $O(n^{1/3}) \approx 100$. Although this is
a significant reduction, it is still greater than the copy number of
MutY ($\sim 20$), so it is likely that the effects of 1D diffusion of
MutY are important. \cite{XIE}

When considering the collective behavior of enzymes, one important
constraint is that the number of BER enzymes available to participate
in the search mechanism is fixed. The copy number for MutY, in
particular, is about 20, \cite{Bai} placing a bound on the total
number of enzymes that can be successfully adsorbed on the DNA
strand. Thus, the CT search mechanism is effective only if the number
of oxoGs is not significantly greater than $\sim 20$.  Although
guanine radicals absorb electrons, thereby seeding the adsorption of
BER enzymes, too many radicals can deplete the reservoir of BER enzyme
before they significantly concentrate to the lesions.

Although in our model, there are two modes of enzyme recruitment --
oxoG-mediated and lesion-mediated (when the lesion is electron
absorbing) -- it is the latter that colocalizes enzymes to
lesions. We re-emphasize that the initial 
recruitment by guanine radicals can only increase
the enzyme density to a level that is of the order of the initial
radical density.  This density is far too low to ensure reliable
excision of the lesion.  However, upon subsequent depositions, enzymes
rapidly colocalize and the accumulation is more focused.

Although our simple model successfully predicts colocalization of CT
BER enzymes to electron-absorbing DNA lesions, it neglects many
potentially important aspects. For example, BER enzymes are not point
particles but have a finite size of about 10-15 base pairs. Random
adsorption of finite sized particles has been studied\cite{DORSOGNA}
and could be used to enhance our current model. We also neglected the
sliding of BER enzymes on DNA.  Inclusion of finite size effects and
enzyme sliding into our model is likely to decrease the search time to
a lesion.  The effect of other proteins on the DNA, besides BER
enzymes, is also important.  These proteins could physically prevent
the adsorption of BER enzymes, absorb electrons emitted by BER enzymes
or shield the lesion from electrons (or possibly all three).  We
currently do not know the effect of molecular crowding on the CT
model, but this topic is discussed by Li \textit{et
al.} \cite{GeneWei} One possible approach to studying these more
subtle attributes is to develop and analyze them within
coarse-grained, mass-action type models, in conjunction with
Monte-Carlo simulations.
\begin{acknowledgements} 
This work was supported by grants from the NSF (DMS-0349195) and the
NIH (K25 AI41935). The authors thank J. Genereux, A. K. Boal and
J. K. Barton for helpful discussions.
\end{acknowledgements}
\appendix
\section{Solution of the one-sided Broadwell problem}
\label{subsec:broadsoln}
Taking the Laplace transform of Eq. (\ref{eqn:master}), we
obtain
\begin{equation}
{\partial  \tilde{{\bf Q}}(x,s) \over \partial x}  =
{\bf M}  \tilde{{\bf Q}}(x,s)
+\left(\begin{array}{c} \delta(x-x_{0}) \\[13pt] 0 \end{array}\right),
\label{EQNPS}
\end{equation}
%
%
where $\tilde{ \bf{Q}}(x,s) = (\tilde{Q}_+(x,s), \tilde{Q}_-(x,s))^T$,
$\tilde{Q}_{\pm}(x,s) \equiv \int_{0}^{\infty}Q_{\pm}(x,t)e^{-s t} \dd t$
%
%
and
\begin{equation}
{\bf M} \equiv \left[\begin{array}{cc}
\displaystyle -(s+\mu+f) & \displaystyle  f \\[13pt]
\displaystyle -f & \displaystyle  s+\mu+f
\end{array}\right].
\label{L}
\end{equation}
The solution to Eq. (\ref{EQNPS}), 
can be found in two separate
regions $x>x_{0}$ and $x<x_{0}$ and matching the solutions with the
appropriate jump conditions derived from integrating Eq. (\ref{EQNPS})
over an infinitessimal segment centered about $x_{0}$:
\begin{equation}
\begin{array}{rl}
\tilde{Q}_+(x_0^+,s) - \tilde{Q}_+(x_0^-,s) &= 1, \\
\tilde{Q}_-(x_0^+,s) - \tilde{Q}_-(x_0^-,s) &= 0.
\end{array}
\label{eqn:JUMP}
\end{equation}
The general solution of
Eq. (\ref{EQNPS}), $\tilde{{\bf Q}}(x,s;x_0,\ell)$, can be expressed in the form
%
\begin{equation}
\tilde{{\bf Q}}(x,s) = \left\{
\begin{array}{ll}
A_{<}\left(\begin{array}{c} 1 \\
c_{1}\end{array}\right)
e^{\lambda_{1}x} + B_{<}\left(\begin{array}{c} 1 \\
c_{2}\end{array}\right)e^{\lambda_{2}x} & , ~x < x_0,\\[14pt]
A_{>}\left(\begin{array}{c} 1 \\
c_{1}\end{array}\right)
e^{\lambda_{1}x} + B_{>}\left(\begin{array}{c} 1 \\
c_{2}\end{array}\right)e^{\lambda_{2}x} & , ~x > x_0,
\end{array} \right.
\label{A4}
\end{equation}
%
%
where $\lambda_{1,2}(s)$, $c_{1,2}(s)$ are given by
\begin{equation}
\begin{array}{l}
\lambda_{1,2}(s) =  \pm \sqrt{(s+\mu)(s+\mu + 2f)}, \\[13pt]
c_{1,2}(s) = \frac{f}{s+\mu+f - \lambda_{1,2}(s)}.
\end{array}
\end{equation}
%

The constants $A_>$, $B_>$, $A_<$, and $B_<$ are obtained
by imposing the Laplace Transformed boundary
conditions $\tilde{Q}_+(0,s) = \tilde{Q}_-(\ell,s) = 0$, which
come from Eq. (\ref{eqn:BC1}), and the jump conditions
(\ref{eqn:JUMP}):
\begin{equation}
\begin{array}{rl}
A_< &= \displaystyle \frac{c_1 c_2 e^{-(\lambda_1 + \lambda_2) x_0}
[ e^{\lambda_1 \ell + \lambda_2 x_0}-e^{\lambda_1 x_0 + \lambda_2 \ell} ]}
{  (c_1 - c_2) ( c_1 e^{\lambda_1 \ell} - c_2 e^{\lambda_2 \ell} )}, \\
& \\
B_< &= \displaystyle \frac{c_1 c_2 e^{-(\lambda_1 + \lambda_2) x_0}
[ e^{\lambda_1 x_0 + \lambda_2 \ell} - e^{\lambda_1 \ell + \lambda_2 x_0}
]}
{  (c_1 - c_2) ( c_1 e^{\lambda_1 \ell} - c_2 e^{\lambda_2 \ell} )},\\
& \\
A_> &= \displaystyle \frac{c_2 e^{\lambda_2 \ell}
[ c_2 e^{-\lambda_1 x_0} - c_1 e^{-\lambda_2 x_0 } ]}
{ (c_1 - c_2) ( c_1 e^{\lambda_1 \ell} - c_2 e^{\lambda_2 \ell} )},\\
& \\
B_> &= \displaystyle \frac{c_1 e^{\lambda_1 \ell}
[ c_1 e^{-\lambda_2 x_0} - c_2 e^{-\lambda_1 x_0} ]}
{(c_1 - c_2) ( c_1 e^{\lambda_1 \ell} - c_2 e^{\lambda_2 \ell} )}.
\end{array}
\end{equation}

\section{Limiting cases of the Broadwell model}
\label{subsec:limit}
Upon eliminating $P_-$ from Eqs. (\ref{EQNP}), $P_+$
satisfies
\begin{equation}
\ptn{P_+}{T}{2} = -2(F+M)\pt{P_+}{T} + V^2 \ptn{P_+}{X}{2} - M^2 P_+.
\label{eqn:mixed}
\end{equation}
%
Similarly, eliminating $P_+$ from Eqs.~(\ref{EQNP}) gives
Eq.~(\ref{eqn:mixed}) but with $P_+$ replaced with $P_-$.  Upon
neglecting electron decay, $M=0$, and Eq.~(\ref{eqn:mixed}) simplifies
to
\begin{equation}
\ptn{P_+}{t}{2} + 2f \pt{P_+}{t} = \ptn{P_+}{x}{2}
\label{eqn:simpleP}
\end{equation}
where we have used the nondimensionalization (\ref{eqn:nondim})
and the non-dimensional flip rate $f = F/(\rho V)$.
When $f \gg 1$, we neglect the first term in
Eq. (\ref{eqn:simpleP}) to obtain a diffusion equation
with diffusivity $1/(2f)$.
When $f \ll 1$, we neglect the second term 
to obtain a wave equation with unit wave speed. These
limits correspond to a diffusive and ballistic electron
motion respectively.

\section{Adiabatic approximation}
\label{subsec:adiabatic}
Since our stochastic analysis does not account for electron-electron
interactions, we assume ``adiabatic'' deposition of BER
enzymes. An adiabatic deposition of enzymes occurs when
each enzyme is deposited sufficiently slowly
so that the emitted electron completes its motion
before the deposition of the next enzyme.  At any given time, there is
at most one traveling electron on the DNA.

\begin{figure}[htbp]
\includegraphics[width=3.4in]{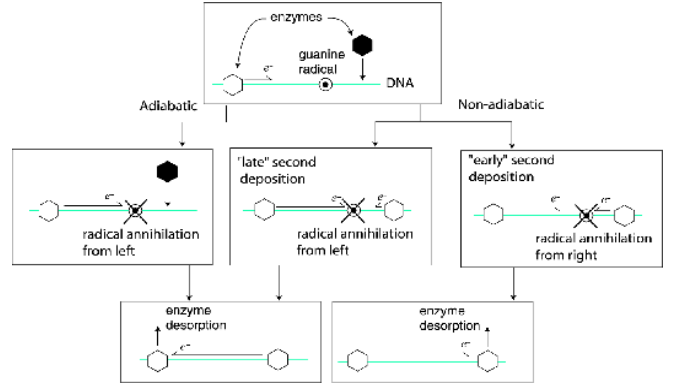}
\caption{
Possible outcomes from an adiabatic and non-adiabatic deposition of a
pair of repair enzymes. The left enzyme is always deposited first, but is
further away from the guanine radical than the right one.  The flip
rate $F$ satisfies $F \ll \rho V$
so that the
electron motion is ballistic. The final configuration of a
non-adiabatic deposition depends critically on the time between the
first and second depositions.}
%
%
\label{fig:nonadiabatic}
\end{figure}

Consider Figure \ref{fig:nonadiabatic}: two enzymes are deposited
on either side of a guanine radical with the left enzyme further away.
For this example, assume that the
electrons are always emitted toward the radical.
In an adiabatic deposition, the deposition of the right enzyme
occurs after the oxoG is annihilated. The final configuration
consists of an adsorbed right enzyme and a desorbed left enzyme.
In a non-adiabatic deposition, the right enzyme can be deposited
before the annihilation of the oxoG. The final enzyme configuration
depends critically on the time between the first and second
depositions. If this time is long (a ``late'' second deposition), the
oxoG is annihilated by the rightward electron and the final
configuration is identical to the adiabatic case. If the
inter-deposition time is short (an ``early'' second deposition), the
leftward electron can annihilate the oxoG first and the final
configuration corresponds to an adsorbed left enzyme and a desorbed
right enzyme.
%

For a deposition to be adiabatic, the electron dynamics must be much
faster than that of enzyme depositions:

\begin{equation}
\rho V, F \gg {k_{\mathrm{on}}\over \rho},
\end{equation}
%
%
where $\rho$ is the density of guanine radicals and $k_{\textrm{on}}$
is an intrinsic enzyme deposition rate per unit length of DNA.
Thus, the adiabatic limit arises when $\rho^{2}V/k_{\mathrm{on}} \rightarrow
\infty$ and $\rho F/k_{\mathrm{on}} \rightarrow \infty$, with
$f=F/(\rho V)$ fixed (to keep the overall probabilities
$\Pi_{\mathrm{adsorb}}, \Pi_{\mathrm{desorb}}$ unchanged in Eq. (\ref{eqn:Qadsorb})).
Note that $f$ can still be small in an adiabatic deposition, 
as is the case in Fig. \ref{fig:nonadiabatic}.
%
%
%

\section{Guanine gap distribution}
\label{subsec:gap}
Consider a lattice made up of $n$ sites on which guanine radicals can
randomly appear at a rate of $\Omega$ radicals per unit time $T$, per
lattice site. Each lattice site can hold at most one guanine radical.
The size of the \emph{gap} between two guanine radicals is the number
of empty sites between them. Let $N(m,T)$ denote the total number of
gaps of size $m$ (measured in lattice sites) at time $T$.  Then
$N(m,T)$ obeys \cite{DORSOGNA}
\begin{equation}
\label{eqn:discrete}
\frac{1}{\Omega} \pt{N(m,T)}{T} = 2 \sum_{m'=m+1}^n N(m',T) - m N(m,T).
\end{equation}
We will take the continuum limit of Eq. (\ref{eqn:discrete}) when the
number of sites becomes infinite, the guanine radicals become points
on a line and the gap length becomes a continuous random variable,
taking any value between 0 and $\infty$.  We aim to calculate the
probability distribution function (PDF) of the gap length given a
fixed average density of guanine radicals $\rho$.

Let $L_0$ be the total length of the lattice and $a$ be the width of a
single lattice site so that $L_0 = n a$. Furthermore, the time taken for $G$ guanines to
appear on the lattice is $T_0$, where $n \Omega T_0 = G$ and $\rho =
G/L_0$.

Now we define dimensionless variables $y$, $t$ and $p = p(y, t)$ where
\begin{eqnarray}
y &=& \rho a m, \label{eqn:y}\\
t &=& T/T_0, \\
p &=& N/(n \Omega T) = N/(Gt).
\label{eqn:p}
\end{eqnarray}
Note that $0 < y < \infty$ and that for large $G$, $Gt$ is
approximately the total number of gaps at time $t$; hence $p$ in
Eq. (\ref{eqn:p}) is the fraction of gaps that have size $N$ at time
$t$. 

The desired continuum limit is now obtained by taking $n \rightarrow
\infty$, $a \rho \rightarrow 0$ so that $y$ in Eq. (\ref{eqn:y})
becomes a continuous variable ranging from $0$ to $\infty$, and $G,L_0
\rightarrow \infty$: the number of radicals that appear and the DNA
length become infinite in such a way that $\rho \equiv G/L_0$ stays a
constant.  When these limits are taken, $p(y,t)$ becomes the
probability of finding a gap of length $y$ at time $t$ and
$\int_0^{\infty} p(y,t) \dd y = 1$.
%
Upon setting $q(y,t) = t p(y,t)$, we obtain the integro-differential
equation
\begin{equation}
\pt{q}{t} = 2\int_y^{\infty} q(y',t)\dd y' - yq(y,t).
\label{eqn:F}
\end{equation}
The  Laplace transform in $t$ of Eq. (\ref{eqn:F}) is
\begin{equation}
s \tilde{q}(y,s) =
2\int_y^{\infty} \tilde{q}(y',s)dy' - y \tilde{q}(y,s), \label{eqn:LT}
\end{equation}
where $\tilde{q}(y,s) = \int_0^{\infty} e^{-st} q(y,t)dt$ and 
we have used the initial condition $p(y,0) = 0$.
Differentiating Eq. (\ref{eqn:LT}) with respect to $y$ gives
\begin{equation}
\frac{d \tilde{q}(y,s)}{dy} + \frac{3 \tilde{q}(y,s)}{(y+s)}  = 0,
\label{eqn:ODE}
\end{equation}
which is solved by $\tilde{q}(y,s) = A(s)/(y+s)^3$, To determine the
integration constant $A(s)$, we take the $y\rightarrow 0$ limit of
Eq. (\ref{eqn:F}) to obtain

\begin{equation}
\left. \pt{q}{t}\right|_{y=0} = 2\int_0^{\infty} q(y',t)\dd y' = 2t, 
\label{eqn:2T}
\end{equation}
where the last equality arises from the normalization of $p(y,t)$.
The Laplace transform of Eq. (\ref{eqn:2T}) gives $\tilde{q}(0,s) =
2/s^3$.  Hence, $\tilde{q}(y,s) = 2/(y+s)^3$ resulting in $q(y,t) =
t^2 e^{-yt}$ and $p(y,t) = t e^{-y t}$.  Therefore, if $Y$ is the non-
dimensionlized gap length at $t=1$, we find
\begin{equation}
\text{Prob}(y \leq Y \leq y + dy) = e^{-y} \dd y. 
\label{eqn:Q}
\end{equation}


%
%
%
%
%

\bibliography{broadwell17b.bbl}

\end{document}